\documentclass[pre,twocolumn,aps,eqsecnum]{revtex4}
\usepackage{epsfig}

\begin{document}

\title{Thermodynamic dislocation theory of high-temperature deformation in aluminum and steel} 
\author{K.C. Le, T.M. Tran} 
\affiliation{Lehrstuhl f\"ur Mechanik - Materialtheorie, Ruhr-Universit\"at Bochum, D-44780 Bochum, Germany}
\author{J.S. Langer}
\affiliation{Department of Physics, University of California, Santa Barbara, California 93106-9530, USA}

\date{\today}

\begin{abstract}
The statistical-thermodynamic dislocation theory developed in previous papers is used here in an analysis of high-temperature deformation of aluminum and steel.  Using physics-based parameters that we expect theoretically to be independent of strain rate and temperature, we are able to fit experimental stress-strain curves for three different strain rates and three different temperatures for each of these two materials.  Our theoretical curves include yielding transitions at zero strain in agreement with experiment. We find that thermal softening effects are important even at the lowest temperatures and smallest strain rates.   
\end{abstract}

\maketitle

\section{Introduction}
\label{Intro} 

Our purpose here is to explore use of the thermodynamic dislocation theory  \cite{LBL-10,JSL-15,JSL-16,JSL-17,JSL-17a} in modelling deformations of materials undergoing thermomechanical processing.  We look at two sets of high-temperature compression tests, one by Shi {\it et al} for aluminum  \cite{SHIetal-97}, and another by Abbod {\it et al} for a steel alloy \cite{ABBODetal-07}.  By making stress-strain measurements over a range of substantially different temperatures and strain rates, and fitting their results to conventional phenomenological formulas, these investigators provided guidance for practical applications in materials processing.  Our question is whether we can do better by using a realistic physics-based theory.  We believe that we can do so and that, in addition, we can obtain basic information about these materials in this way.  

Our new ability to interpret data of the kind published in  \cite{SHIetal-97} and  \cite{ABBODetal-07} is a result of the fact that, in its latest versions  \cite{JSL-17,JSL-17a}, the thermodynamic dislocation theory includes a description of yielding transitions.  Earlier versions of the theory were based on data for copper as shown, for example, in Kocks and Mecking \cite{KOCKS-MECKING-03} or in Meyers {\it et al.} \cite{MEYERSetal-95}.  There, the onset of hardening occurs at a negligibly small stress corresponding to a negligibly small density of dislocations, so that one of the central parameters in the theory can be obtained directly from experiment.  Most stress-strain curves in the literature such as the ones to be studied here exhibit nonzero yield stresses near effectively zero strain.  With the present theory and with experimental data of the kind to be used here, these situations now can be studied systematically.  As will be seen, however, the fact that these experiments were not carried out with a physics-based theory in mind makes their interpretation problematic at some places. 

The thermodynamic dislocation theory is based on two unconventional ideas.  The first of these is that, under nonequilibrium conditions, the atomically slow configurational degrees of freedom of deforming solids are characterized by an effective disorder temperature that differs from the ordinary thermal temperature.  Both of these temperatures are thermodynamically well defined variables whose equations of motion determine the irreversible behaviors of these systems.  The second  principal idea is that entanglement of dislocations is the overwhelmingly dominant cause of resistance to deformation in polycrystalline materials.  These two ideas have led to successfully predictive theories of strain hardening \cite{LBL-10,JSL-15}, steady-state stresses over exceedingly wide ranges of strain rates \cite{LBL-10}, adiabatic shear banding \cite{JSL-16,JSL-17}, and Hall-Petch effects \cite{JSL-17a}.  

We start in Sec.~\ref{EOM} with a brief annotated summary of the equations of motion to be used here.  Our focus is on the physical significance of the various parameters that occur in them.  We discuss which of these parameters are expected to be material-specific constants, independent of temperature and strain rate, and thus to be key ingredients of the theory.  Shi {\it et al.}  \cite{SHIetal-97} and Abbod {\it et al} \cite{ABBODetal-07} each provide nine different stress-strain curves, for three temperatures and three strain rates, for aluminum and steel respectively.  As will be seen, this is enough data for us to use in constructing theories, but these data sets are not immune to experimental uncertainties.

In Sec.~\ref{DATA1}, we show the data for pure aluminum \cite{SHIetal-97}, describe our methods for using that data to determine the material-specific parameters, and describe our theoretical interpretation of those measurements.  These analyses are extended to the steel data \cite{ABBODetal-07} in Sec.~\ref{DATA2}.  We conclude in Sec.\ref{CONCLUSIONS} with some remarks about the significance of these calculations.
 
\section{Equations of Motion}
\label{EOM}

Strictly speaking, the thermodynamic dislocation theory should be written in three-dimensional tensorial notation in order to use it in analyses  of plane-strain  compression tests. There is no fundamental reason why this cannot be done.  For example, Rycroft and Bouchbinder \cite{RYCROFT-EB-12,RYCROFT-EB-16} used a simple tensorial version of the shear-transformation-zone theory \cite{FL-11}  in their moving-boundary analysis of fracture toughness in metallic glasses.   Moreover, Fig.1 in \cite{SHIetal-97a} shows a diagram of a plane-strain sample like those used in \cite{SHIetal-97}. Here, a thin rectangular block  under uniaxial compression is shown bulging at its sides and thinning at its center in addition to undergoing pure shear.  These deformations if actually as large as shown would slightly affect our interpretation of the reported stress-strain data.  However, a detailed analysis of those deformations would be well beyond the scope and needs of this project.  

Suppose, for simplicity, that the experimental sample is a two dimensional  rectangular block in the $x\,y$ plane, being compressed between two rigid plates parallel to the $x$ axis. The compressive stress in the $y$ direction is $\sigma _{yy} \equiv -\sigma $. If the plates are well lubricated so that the friction between the block and the plates is negligible, then $\sigma _{xx}\approx 0$. For this uniaxial geometry, the stress tensor is naturally expressed in the $x'\,y'$ frame of reference oriented at 45$^\circ $ to the $x\,y$ axes. In that frame, the shear stress is $\sigma _{x'y'} = \sigma_{y'x'}= -\sigma /2$. If the material  is incompressible, then the total elastic plus plastic strain rate is $\dot{\epsilon }_{yy}=-\dot{\epsilon}_{xx}\equiv -\dot{\epsilon}$.  In the rotated frame, the shear rate is $\dot\epsilon_{x'y'} = \dot\epsilon_{y'x'} = - \dot\epsilon$.  As usual, we assume that the elastic and plastic strain rates are simply additive, e.g. $\dot\epsilon_{x'y'} = \dot\epsilon^{el}_{x'y'}+\dot\epsilon^{pl}_{x'y'}$ .  Then, by convention, we write $\dot\sigma_{x'y'}= 2\,\mu\,\dot\epsilon^{el}_{x'y'}$, where $\mu$ is the shear modulus and the factor $2$ accounts for the distinction between ``true'' and  ``engineering'' strain. Putting these pieces together, we write $\dot{\sigma }=\alpha\, \mu \,(\dot{\epsilon }-\dot{\epsilon }^{pl})$, where  $\alpha \simeq 4$ is a geometric factor, and we have dropped the directional subscripts. In this way we have recovered the one-dimensional notation used in earlier papers in this series and in much of the literature in this field.  

Now assume that this spatially uniform system is driven at a constant shear rate $\dot\epsilon \equiv Q/\tau_0$, where $\tau_0 \equiv 10^{-12} s$ is a characteristic microscopic time scale. This motion is driven by the time dependent shear stress $\sigma$. Because the system is undergoing steady-state shear, we can replace the time $t$ by the total strain $\epsilon$ so that $\tau_0\,\partial/\partial t \to Q\,\partial/\partial \epsilon$. Then denote the dimensionless plastic strain rate by $q(\epsilon) \equiv \tau_0\,\dot\epsilon^{pl}(\epsilon)$. The equation of motion for the stress becomes
\begin{equation}
\label{sigmaeqn}
{d\sigma\over d \epsilon} = \alpha\, \mu\,\left[1 - {q(\epsilon)\over Q}\right].
\end{equation}

The internal state variables that describe this system are the areal density of dislocations $\rho\equiv  \tilde\rho/ b^2$ (where $b$ is the length of the Burgers vector), the effective temperature $\tilde\chi$ (in units of a characteristic dislocation energy, say $e_D$), and the ordinary temperature $\tilde\theta$ (in units of the pinning temperature $T_P = e_P/k_B$, where $e_P$ is the pinning energy defined below).  Note that  $b/\sqrt{\tilde\rho}$ is the average distance between dislocations. All three of these dimensionless quantities, $\tilde\rho$, $\tilde\chi$, and $\tilde\theta$, are functions of $\epsilon$. 

The central, dislocation-specific ingredient of this analysis is the thermally activated depinning formula for the dimensionless plastic strain rate $q$ as a function of a non-negative stress $\sigma$:  
\begin{equation}
\label{qdef}
q(\epsilon) = \sqrt{\tilde\rho} \,\exp\,\Bigl[-\,{1\over \tilde\theta}\,e^{-\sigma/\sigma_T(\tilde\rho)}\Bigr]. 
\end{equation}
This is an Orowan relation of the form $q = \rho\,b\,v\,\tau_0$ in which the speed of the dislocations $v$ is given by the distance between them multiplied by the rate at which they are depinned from each other.  That rate is approximated here by the activation term in Eq.(\ref{qdef}), in which the energy barrier $e_P$ (implicit in the scaling of $\tilde\theta$) is reduced by the stress dependent factor $e^{-\sigma/\sigma_T}$, where  $\sigma_T(\tilde\rho)= \mu_T\,\sqrt{\tilde\rho}$ is the Taylor stress, and $\mu_T \equiv r\, \mu$.  The dimensionless number $r$ is the ratio of a depinning length to the length of the Burgers vector, for convenience divided here by the geometrical factor $\alpha$ associated with the stress $\sigma$.  Thus, $r$ should be approximately independent of temperature and strain rate.  Note that only the magnitude of $\sigma$ appears in this expression for a local time scale.  Directional information would be included in tensorial equations of motion for stress fields and flow patterns, but not in this expression for a scalar time scale.  

The pinning energy $e_P$ is large, of the order of electron volts, so that $\tilde\theta$ is very small.  As a result, $q(\epsilon)$ is an extremely rapidly varying function of $\sigma$ and $\tilde\theta$.  This strongly nonlinear behavior is the key to understanding yielding transitions and shear banding as well as many other important features of polycystalline plasticity.  For example, the extremely slow variation of the steady-state stress as a function of strain rate discussed in  \cite{LBL-10} is the converse of the extremely rapid variation of $q$ as a function of $\sigma$ in Eq.(\ref{qdef}). In what follows, we shall see that this temperature sensitivity of the strain rate is the key to understanding important aspects of the thermomechanical behavior.  

The equation of motion for the scaled dislocation density $\tilde\rho$ describes energy flow. It says that some fraction of the power delivered to the system by external driving is converted into the energy of dislocations, and that that energy is dissipated according to a detailed-balance analysis involving the effective temperature $\tilde\chi$.  This equation is: 
\begin{equation}
\label{rhodot}
{\partial\tilde\rho\over \partial\epsilon} = \kappa_1\,{\sigma\,q\over \nu(\tilde\theta,\tilde\rho,Q)^2\,\mu_T\,Q}\, \Bigl[1 - {\tilde\rho\over \tilde\rho_{ss}(\tilde\chi)}\Bigr],
\end{equation}
where $\tilde\rho_{ss}(\tilde\chi) = e^{- 1/\tilde\chi}$ is the steady-state value of $\tilde\rho$ at given $\tilde\chi$.  The coefficient $\kappa_1$ is an energy conversion factor that, according to arguments presented in  \cite{LBL-10} and \cite{JSL-17}, should be independent of both strain rate and temperature.  The other quantity that appears in the prefactor in Eq.(\ref{rhodot}) is
\begin{equation}
\label{nudef}
\nu(\tilde\theta,\tilde\rho,Q) \equiv \ln\Bigl({1\over \tilde\theta}\Bigr) - \ln\Bigl[\ln\Bigl({\sqrt{\tilde\rho}\over Q}\Bigr)\Bigr].
\end{equation}

\begin{figure}[h]
\centering \epsfig{width=.5\textwidth,file=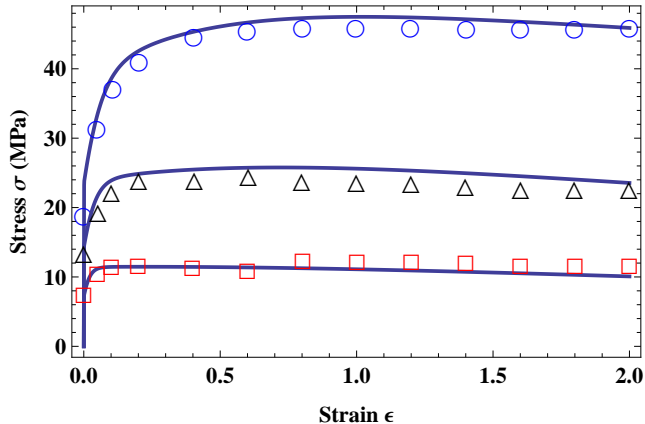} \caption{(Color online) Stress-strain curves for aluminum at the small strain rate $\dot\epsilon = 0.25\,s^{-1}$, for temperatures $300\,C,~400\,C,~500\,C$  shown from top to bottom.  The experimental points are taken from Shi {\it et al.} \cite{SHIetal-97}} \label{Al-Fig-1}
\end{figure}

\begin{figure}[h]
\centering \epsfig{width=.5\textwidth,file=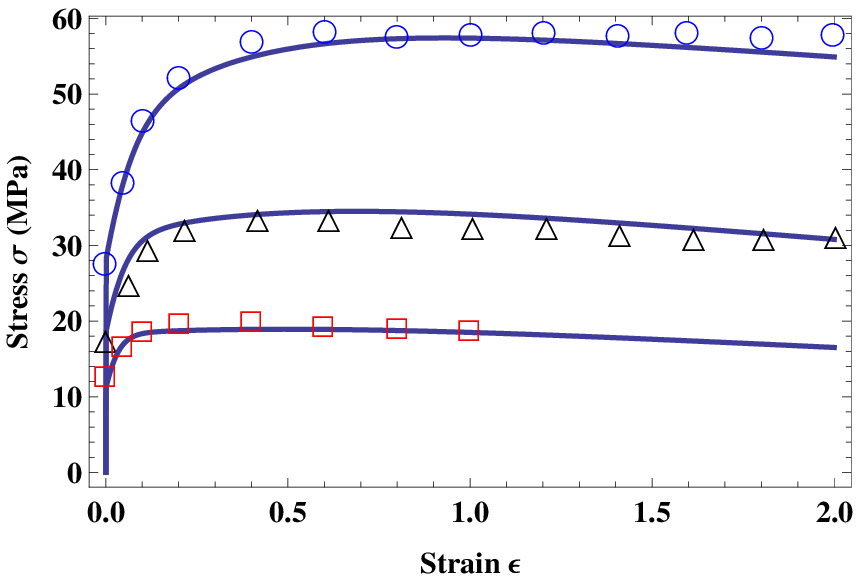} \caption{(Color online) Stress-strain curves for aluminum at the strain rate $\dot\epsilon = 2.5\,s^{-1}$, for temperatures $300\,C,~400\,C,~500\,C$ shown from top to bottom.  The experimental points are taken from Shi {\it et al.} \cite{SHIetal-97}} \label{Al-Fig-2}
\end{figure}

The equation of motion for the scaled effective temperature $\tilde\chi$ is a statement of the first law of thermodynamics for the configurational subsystem: 
\begin{equation}
\label{chidot}
{\partial\,\tilde\chi\over \partial\epsilon} = \kappa_2\,{\sigma\,q\over \mu_T\,Q}\,\Bigl( 1 - {\tilde\chi\over \tilde\chi_0} \Bigr). 
\end{equation}
Here, $\tilde\chi_0$ is the steady-state value of $\tilde\chi$ for strain rates appreciably smaller than inverse atomic relaxation times, i.e. much smaller than $\tau_0^{-1}$. The dimensionless factor $\kappa_2$ is inversely proportional to the effective specific heat $c_{e\!f\!f}$. Unlike $\kappa_1$, there is no  reason to believe that $\kappa_2$ is a rate-independent constant.  In \cite{JSL-17a}, $\kappa_2$ for copper was found to decrease from $17$ to $12$ when the strain rate increased by a factor of $10^6$.  Since we shall consider changes in strain rate of at most a factor of $10^2$ here, we shall assume that $\kappa_2$ is a constant.

\begin{figure}[h]
\centering \epsfig{width=.5\textwidth,file=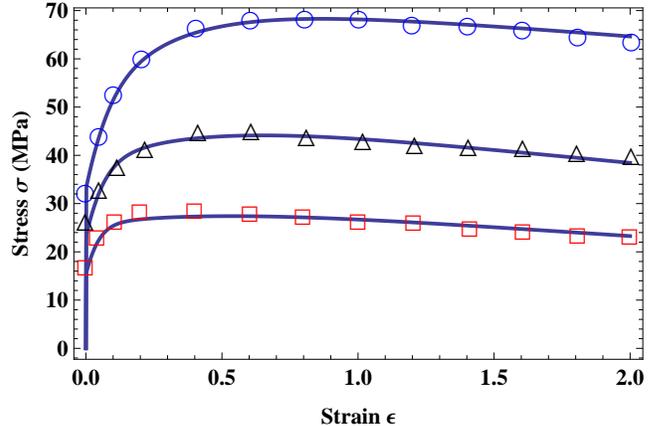} \caption{(Color online) Stress-strain curves for aluminum at the highest strain rate $\dot\epsilon = 25\,s^{-1}$, for temperatures $300\,C,~400\,C,~500\,C$ shown from top to bottom.  The experimental points are taken from Shi {\it et al} \cite{SHIetal-97}} \label{Al-Fig-3}
\end{figure}

The equation of motion for the scaled, ordinary temperature $\tilde\theta$ is 
\begin{equation}
\label{thetadot}
{\partial\tilde\theta\over \partial\epsilon} = K(\tilde\theta)\,{\sigma\,q\over Q} - {K_2\over Q}\,(\tilde\theta - \tilde\theta_0).
\end{equation} 
Here, $K(\tilde\theta) = \beta/ (T_P\,c_p\,\rho_d)$ is a thermal energy conversion factor.   $c_p$ is the thermal heat capacity per unit mass, $\rho_d$ is the mass density, and $0< \beta < 1$ is a dimensionless constant known as the Taylor-Quinney factor. As indicated here,  $K(\tilde\theta)$ will be found to be non-trivially temperature dependent for both of the materials discussed in Secs.~\ref{DATA1} and \ref{DATA2}.  $K_2$ is a thermal transport coefficient that controls how rapidly the system relaxes toward the ambient temperature $T_0$, that is, $\tilde\theta \to \tilde\theta_0 = T_0/T_P$.  This coefficient turns out to be too small to be measured for the situations reported here; but that will not always be the case.  In principle, after long enough times of steady deformation, systems must reach steady-state temperatures determined by the balance between heating and cooling terms in Eq.~(\ref{thetadot}).   

\section{Data Analysis: Aluminum}
\label{DATA1}

The experimental results of Shi {\it et al.}  \cite{SHIetal-97} for aluminum, along with our theoretical results based on the equations of motion in Sec.~\ref{EOM}, are shown in Figs. \ref{Al-Fig-1}, \ref{Al-Fig-2}, and \ref{Al-Fig-3}. These figures are presented in order of increasing strain rate, $\dot\epsilon = 0.25\,s^{-1},~2.5\,s^{-1}$, and $25\,s^{-1}$. Within each figure are curves for the three different temperatures $300\,C,~400\,C$, and $~500\,C$ (blue circles, black triangles, and red squares respectively) shown from top to bottom. 

In order to compute the theoretical curves in these figures, we need values for five system-specific parameters: the activation temperature $T_P$, the stress ratio $r$, the steady-state scaled effective temperature $\tilde\chi_0$, and the two dimensionless conversion factors $\kappa_1$ and $\kappa_2$.  We also need initial values of the scaled dislocation density $\tilde\rho(\epsilon = 0) \equiv \tilde\rho_i$ and the effective temperature $\tilde\chi(\epsilon = 0) \equiv \tilde\chi_i$, which are determined by sample preparation -- presumably the same for all samples, but possibly a source of experimental uncertainty.  In addition, we need a formula for the thermal conversion factor $K(\tilde\theta)$ in Eq.~(\ref{thetadot}) which, for aluminum, we can take to have the linear form  
\begin{equation}
\label{Ktheta}
K(\tilde\theta) = K_0 \,\left[1 + c_1\,T_P\,(\tilde\theta - \tilde\theta_1)\right],
\end{equation}
where $T_P\,\tilde\theta_1$ is a reference temperature, chosen here to be $573\,K$.  The numbers $K_0$ and $c_1$ remain to be determined from the data.  Finally, we need a formula for the temperature dependent shear modulus $\mu(T)$, which we take from  \cite{VARSHNI-70,CHENetal-98} to be 
\begin{equation}
\label{muAl}
\mu(\tilde\theta) = \mu_1 - \left[{D\over \exp(T_1/T_P\,\tilde\theta)-1}\right],
\end{equation}
where $\mu_1 = 28.8 \, GPa$, $D = 3.44\,GPa$, and $T_1 = 215\,K$. (A simple linear approximation to this formula analogous to Eq.~(\ref{Ktheta}) would be completely adequate for our purposes.)  

In earlier papers starting with \cite{LBL-10}, we were able to begin evaluating the parameters by observing steady-state stresses $\sigma_{ss}$ at just a few strain rates $Q$ and ambient temperatures $T_0 = T_P\,\tilde\theta_0$, and inverting Eq.~(\ref{qdef}) to find 
\begin{equation}
\label{sigmass}
\sigma_{ss} = r\,\mu\,\sqrt{\tilde\rho_{ss}}\,\,\nu(\tilde\theta_0,\tilde\rho_{ss},Q);~~~~\tilde\rho_{ss} = e^{- 1/\tilde\chi_0}.
\end{equation}
Knowing $\sigma_{ss}$, $T_0$ and $Q$ for three stress-strain curves, we could solve this equation for $T_P$, $r$, and $\tilde\chi_0$, and check for consistency by looking at other steady-state situations. With that information, it was relatively easy to evaluate $\kappa_1$ and $\kappa_2$ by directly fitting the full stress-strain curves.  This strategy does not work here because the thermal effects are highly nontrivial.  Examination of the experimental data shown in the figures indicates that almost all of these samples are undergoing thermal softening at large strains; the stresses are decreasing and the temperatures must be increasing.  Even the curves that appear to have reached some kind of steady state have not, in fact, done so at their nominal ambient temperatures.  

\begin{figure}[h]
\centering \epsfig{width=.5\textwidth,file=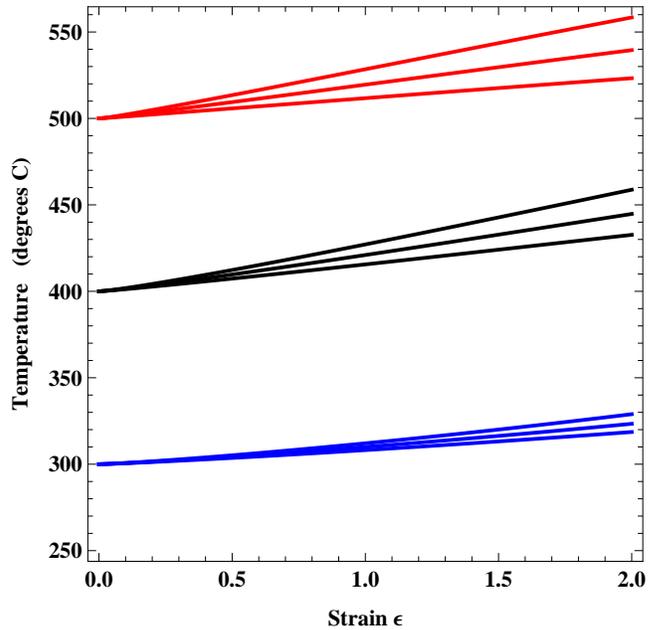} \caption{(Color online) Temperature as a function of strain for each of the nine stress-strain tests shown for aluminum in the preceding figures. The initial ambient temperatures are $300\,C,\, 400\,C$ and $500\,C$ (blue, black, and red) as seen on the left axis.  Each group of three curves is for strain rates of $\dot\epsilon = 0.25\,s^{-1},~2.5\,s^{-1}$, and $25\,s^{-1}$, from bottom to top. } \label{Al-Fig-4}
\end{figure}

To counter this difficulty, we have resorted to large-scale least-squares analyses. (A preliminary discussion of this procedure has been presented by two of us, Le and Tran \cite{LE-TRAN-17}.) That is, we have computed the sum of the squares of the differences between our theoretical stress-strain curves and the experimental points, and have minimized this sum in the space of the unknown parameters.  We have explored options of omitting some of the data, fitting the theory to just those portions of the data that seemed most reliable.  For example, we have looked to see what happens if we omit the yield points in this calculation on the assumption that they are most sensitive to variations in sample preparation.  Our results appear to be robust.  We find: $T_P = 2.40\times 10^4\,K,~r = 0.040,~\chi_0 = 0.249,~\kappa_1=0.97,~\kappa_2= 12,~ \tilde\rho_i = 0.0035,\,\tilde\chi_i = 0.224,~ K_0 = 7.0 \times 10^{-6},~ c_1 = 0.0257$, and $K_2 = 0$.  So far as we can tell, our values of $K_0$ and $c_1$ are consistent with values of the Taylor-Quinney factor $\beta$ of the order of unity or less.  For simplicity, we have set $\alpha = 1$ in Eq.~(\ref{sigmaeqn}) because the  slopes of the initial elastic parts of the stress-strain curves are too large to be meaningful here.  Note, however, that with $\alpha \cong 4$ and $r = 0.04$, the ratio of the depinning length to the length of the Burgers vector becomes $0.16$, which seems physically reasonable.  

The agreement between theory and experiment seems to us to be well within the bounds of experimental uncertainties.  Even the initial yielding transitions appear to be described accurately by this dynamical theory. There are only a few visible discrepancies.  For example, the experimental data in Fig. \ref{Al-Fig-1} for $\dot\epsilon = 0.25\,s^{-1},~ T = 500\,C$ exhibit a small, abrupt increase in the stress at about $\epsilon \cong 0.8$, which may indicate some kind of instrumental problem.  Also, the stresses for $T = 400\,C$ in that figure are slightly below those predicted by the theory, and there is a smaller discrepancy of the opposite sign on the curve at $\dot\epsilon = 2.5\,s^{-1},~ T = 300\,C$ in Fig. \ref{Al-Fig-2}.  Nothing about these results leads us to believe that there are relevant physical ingredients missing in the theory. 

\begin{figure}[h]
\centering \epsfig{width=.5\textwidth,file=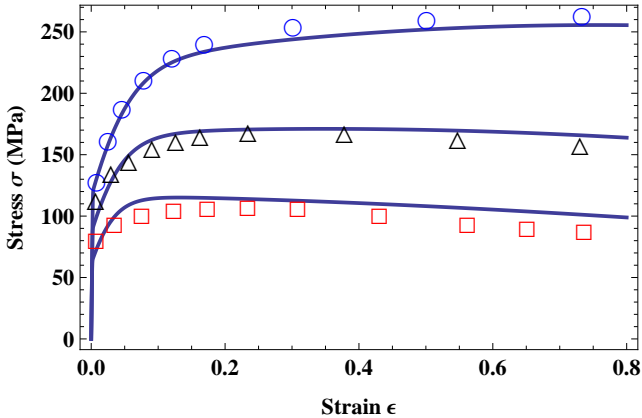} \caption{(Color online) Stress-strain curves for steel at the small strain rate $\dot\epsilon = 0.1\,s^{-1}$, for temperatures $850\,C,~950\,C,~1050\,C$ shown from top to bottom.  The experimental points are taken from Abbod {\it et al} \cite{ABBODetal-07}} \label{Steel-Fig-5}
\end{figure}

\begin{figure}[h]
\centering \epsfig{width=.5\textwidth,file=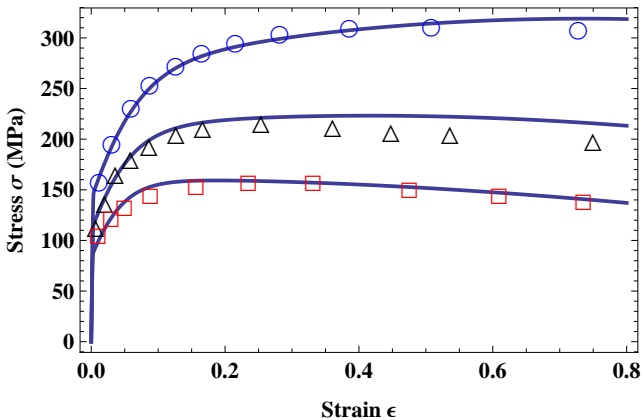} \caption{(Color online) Stress-strain curves for steel at the strain rate $\dot\epsilon = 1.0\,s^{-1}$, for temperatures $850\,C,~950\,C,~1050\,C$ shown from top to bottom.  The experimental points are taken from Abbod {\it et al} \cite{ABBODetal-07}} \label{Steel-Fig-6}
\end{figure}

To complete our analysis of the Shi {\it et al.} data for pure aluminum, we show in Fig. \ref{Al-Fig-4}  our computed temperatures as functions of strain  for each of the nine stress-strain curves shown in the preceding figures.  Here, we may be finding an interesting discrepancy between our interpretation and that of Shi {\it et al.}.  Those authors say that ``In the high strain rate tests, particularly at low temperatures, temperature rises of up to $30\, K$ were observed at the start of steady state."  We do see temperature rises of roughly that magnitude.  However, as stated above, we do not think that these tests have reached steady state, especially not the one at the highest strain rate and lowest temperature shown at the top of Fig.~\ref{Al-Fig-3}, which clearly is still softening at large strain.  Also, as shown in Fig.~\ref{Al-Fig-4}, we predict that the larger temperature increases occur at the higher ambient temperatures because our data analysis tells us that the thermal conversion factor $K(\tilde\theta)$ in Eq.~(\ref{thetadot}) is larger there. 

\begin{figure}[h]
\centering \epsfig{width=.5\textwidth,file=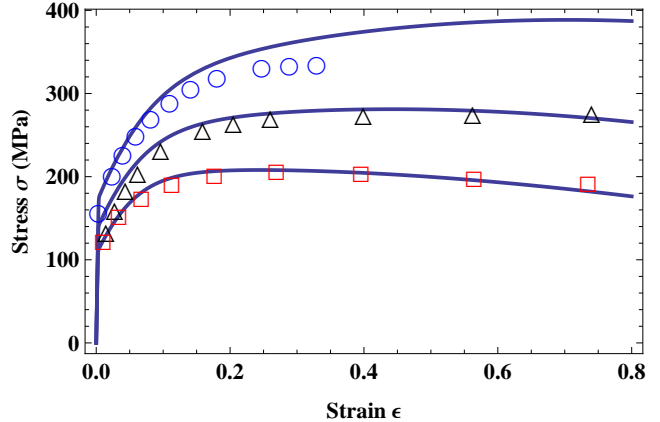} \caption{(Color online) Stress-strain curves for steel at the highest strain rate $\dot\epsilon = 10\,s^{-1}$, for temperatures $850\,C,~950\,C,~1050\,C$ shown from top to bottom.  The experimental points are taken from Abbod {\it et al} \cite{ABBODetal-07}} \label{Steel-Fig-7}
\end{figure}

Shi {\it et al.} \cite{SHIetal-97} also show stress-strain curves for aluminum alloys Al-1\%Mn and Al-1\%Mg.  We have tried to analyze these data sets using the same techniques that we used for pure aluminum but have concluded that this is not a useful exercise.  The main problem is that the experimental results show anomalously increasing stresses at large strains, which Shi {\it et al.} attribute to breakdowns of the lubrication layers between the samples and their instrumental supports.  We have tried to guess which portions of the experimental curves might be unaffected by the lubrication problem; but we have not succeeded in obtaining plausible, self-consistent results.

\section{Data Analysis: Steel}
\label{DATA2}

As a second example of thermal processing data, we turn to the Fe-30\% Ni austenitic alloy studied by Abbod {\it et al} \cite{ABBODetal-07}.  According to those authors, this alloy is a good model material for studying hot deformation of the austenitic phases of carbon-manganese steels.  For simplicity, we refer to it henceforth simply as ``steel.'' We have digitized the experimental data from their Fig. 1 and show it here in Figs.~\ref{Steel-Fig-5}, \ref{Steel-Fig-6} and \ref{Steel-Fig-7}.  In analogy to our presentation of the aluminum data in Sec.~\ref{DATA1}, these figures are shown in order of increasing strain rate, $\dot\epsilon = 0.1\,s^{-1},~1.0\,s^{-1}$, and $10\,s^{-1}$. Within each figure are curves for the three different temperatures $850\,C,~950\,C$, and $~1050\,C$ (blue circles, black triangles, and red squares respectively) shown from top to bottom.

\begin{figure}[h]
\centering \epsfig{width=.5\textwidth,file=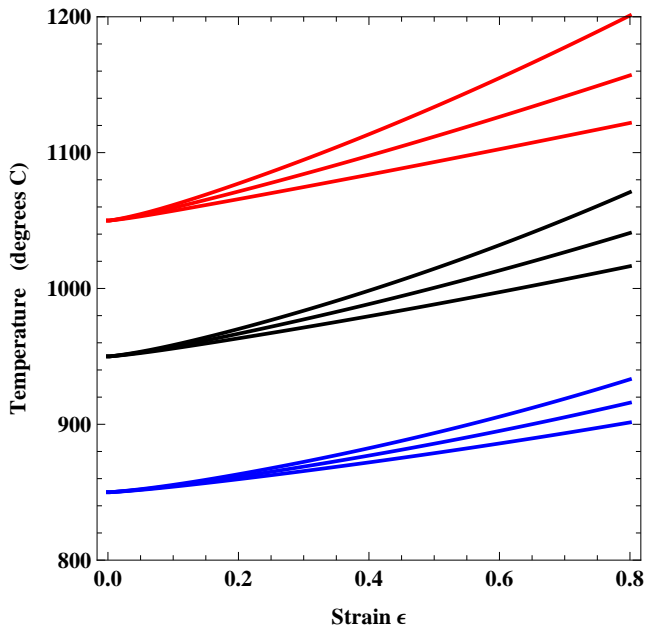} \caption{(Color online) Temperature as a function of strain for each of the nine stress-strain tests shown for steel in the preceding figures. The initial ambient temperatures are $850\,C,\, 950\,C$ and $1050\,C$ (blue, black, and red) as seen on the left axis.  Each group of three curves is for strain rates of $\dot\epsilon = 0.10\,s^{-1},~1.0\,s^{-1}$, and $10\,s^{-1}$, from bottom to top. } \label{Steel-Fig-8}
\end{figure}

In analyzing this data, we have used the same least-squares method that we used for aluminum.  We find: $T_P = 4.59\times 10^4\,K,~r = 0.122,~\chi_0 = 0.284,~\kappa_1=0.958,~\kappa_2= 5.43,~ \tilde\rho_i = 0.0023,\,\tilde\chi_i = 0.215$, and $K_2 = 0$. The one interesting difference is that a slightly nonlinear thermal conversion factor of the form
\begin{equation}
\label{Ktheta*}
K(\tilde\theta) = K^*\,e^{- T^*/T_P\,\tilde\theta}
\end{equation}
seems to produce a better fit to the data than the linear form used previously.  We find $K^* = .00879$ and $T^* = 8390\,K$.  The activated form of this equation is suggestive but probably not meaningful; note that we use it only over a narrow range of temperatures. We also use the following approximation for the shear modulus (derived from data given in \cite{NAVA-13}):
\begin{equation}
\label{muSteel}
\mu(\tilde\theta) = 85,970 - 33.6\,T_P\,\tilde\theta + 0.0009\, (T_P\,\tilde\theta)^2.
\end{equation}

Once again, the results of this analysis seem to be within the bounds of experimental uncertainties.  The one visible discrepancy is for the top curve in Fig.~\ref{Steel-Fig-5}, for $\dot\epsilon = 10\,s^{-1}$ and ambient temperature $850\,C$, where the experimental data drops below our prediction at a relatively small strain.  

The potentially most serious discrepancy pertains to the strain dependence of our predicted temperatures,  shown here in Fig.~\ref{Steel-Fig-8} in analogy to the temperatures for aluminum shown in Fig.~\ref{Al-Fig-4}.  Supposedly the same temperatures are shown by Abbod {\it et al} \cite{ABBODetal-07} in their Fig.2; but those temperatures are not measured directly.  Apparently they are computed from the stress-strain data, perhaps using a temperature-independent thermal conversion factor. Their orders of magnitude and growth as functions of strain rate at fixed ambient temperatures are similar to our results; but their dependence on the ambient temperatures themselves is qualitatively different. 

Note finally that, with $r = 0.122$ and $\alpha = 4$, the ratio of the depinning length to the length of the Burgers vector becomes $0.48$ which, if true, would imply an interestingly nontrivial atomic-scale structure for the interaction between dislocations.   

\section{Concluding Remarks}
\label{CONCLUSIONS}

On the whole, these results seem to us to be quite satisfactory.  Note that we now are using the thermodynamic dislocation theory not just to test its validity but also as a tool for discovering properties of structural materials.  For example, we did not know at the beginning of this investigation that thermal softening would play so important a role even for the samples subjected to very slow deformations at moderately low temperatures.  One of the main reasons for the success of this theory -- as has been emphasized here and in earlier papers -- is the extreme sensitivity of the plastic strain rate to small changes in the temperature or the stress.

To put this point in perspective, note the difference between the expression for the dimensionless plastic strain rate $q$ in Eq.~(\ref{qdef}) and the phenomenological approach adopted by Shi {\it et al.} and Abbod {\it et al}.  Both of these groups of investigators base their analyses on the Zener-Hollomon parameter which, in the present notation, is $Z \equiv \dot\epsilon\, \exp\,(T_Z/T_P\,\tilde\theta)$, where $T_Z$ is an activation temperature analogous to $T_P$.   They express their results for different stresses, strains, strain rates and temperatures as functions of $Z$ which, in analogy to Eq.~(\ref{qdef}), means that their strain rate  $\dot\epsilon$ is proportional to the activation factor $\exp\,(-\,T_Z/T_P\,\tilde\theta)$ multiplied by some function of the stress.  By fitting their data in this way, they find $T_Z/T_P \cong 0.79$ for aluminum and $1.7$ for steel.  In other words, their estimated activation energies are of roughly the same magnitude as ours.  

One crucial difference between our approach and theirs is that, in Eq.~(\ref{qdef}), the depinning activation barrier is itself a function of the stress and the dislocation density.  In this way, the thermodynamic dislocation theory is qualitatively different from conventional theories dating back to Peierls and Nabarro in which dislocations are perceived to be gliding independently through imperfect lattices, resisted by barriers whose dynamical properties are independent of the dislocations themselves.  That is not what is happening in the thermodynamic dislocation theory.  The nonlinear sensitivity to thermal variations that appears in the present investigation is just a mild version of the same dynamical mechanism that produces yielding transitions and adiabatic shear bands, which have been beyond the reach of conventional dislocation theories.  

Even more importantly, the conventional theories are not truly dynamic.  For example, in a fully dynamic theory, an activation factor such as the one occurring in the Zener-Hollomon formula should mean that an increase in temperature produces an increase in strain rate which, in turn, increases the rate of heat generation. This is the nonlinear feedback loop that produces the thermal softening seen in this paper and the runaway instability in the theory of adiabatic shear banding \cite{JSL-17}. But it is not easy to see how such an equation of motion could be incorporated into conventional phenomenological descriptions of dislocation enabled plasticity. We believe that we have found better ways to make progress in this field by focussing on the nonequilibrium statistical thermodynamics of these systems.  

\begin{acknowledgments}

T.M. Tran acknowleges support from the Vietnamese Government Scholarship Program 911.  JSL was supported in part by the U.S. Department of Energy, Office of Basic Energy Sciences, Materials Science and Engineering Division, DE-AC05-00OR-22725, through a subcontract from Oak Ridge National Laboratory.

\end{acknowledgments}

\end{document}